\documentstyle[preprint,aps,floats,epsfig]{revtex}

\def\real{I\negthinspace R}

\def\sech{\,{\rm sech}}
\newcommand{\be}{\begin{equation}}
\newcommand{\ee}{\end{equation}}
\newcommand{\bea}{\begin{eqnarray}}
\newcommand{\eea}{\end{eqnarray}}
\newcommand{\bml}{\begin{mathletters}}
\newcommand{\eml}{\end{mathletters}}

\begin{document}
\preprint{DTP/99/57, gr-qc/0002063}
\draft
\tighten

\title{The Dynamics of Curved Gravitating Walls}
\author{Filipe Bonjour$^1$,%
        \footnote{E-mail address: \texttt{Bonjour@ime.unicamp.br}}
        Christos Charmousis$^2$,\footnote{E-mail address:
        \texttt{Christos.Charmousis@durham.ac.uk}} and
        Ruth Gregory$^2$\footnote{E-mail address:
        \texttt{R.A.W.Gregory@durham.ac.uk}}}
\address{~$^1$ Departamento de Matem\'etica Aplicada, IMECC,
         Universidade Estadual de Campinas, \\ CP 6065, 13081-970 Campinas SP,
         Brazil.\\
         ~$^2$ Centre for Particle Theory, 
         Durham University, South Road, Durham, DH1 3LE, U.K.}
\date{\today}
\setlength{\footnotesep}{0.5\footnotesep}
\maketitle

\begin{abstract}
We examine the dynamics of a self-gravitating domain wall using the $\lambda
\Phi^4$ model as a specific example. We find that the Nambu motion of the
wall is quite generic and dominates the wall motion even in the presence of
gravity. We calculate the corrections to this leading order motion, and
estimate the effect of the inclusion of gravity on the dynamics of the wall.
We then treat the case of a spherical gravitating thick wall as a particular
example, solving the field equations and calculating the corrections to the
Nambu motion analytically for this specific case. We find that the presence
of gravity retards collapse in this case.
\end{abstract}

\pacs{PACS numbers: 98.80.Cq, 11.27.+d, 04.40.-b\hfill gr-qc/0002063}


%
%

\section{Introduction}

Over the past few decades, topological defects have become a familiar class of
objects in many areas of physics. In cosmology, defects are believed to arise
generically during phase transitions in the early Universe, and have been
notably invoked to account for the anisotropies which seeded cosmological
structures~\cite{VS}. Despite the discouraging discovery that the power
spectrum of global strings does not agree with observations of the COBE
satellite~\cite{PoSp}, recent claims of a nongaussian component in the
microwave background~\cite{NonGauss} (see also~\cite{NGN}), and the improved
agreement of the spectrum predictions for models of cosmic strings with a
cosmological constant~\cite{StrLam} (whose best fit coincides with the value of
$\Lambda$ determined from type~Ia supernovae~\cite{SN}) suggest that it may be
still too early to discard them as the source of cosmic structure.

Domain walls are defects that arise when the phase transition occurs by the
breakdown of a discrete symmetry. They correspond to solitons in $1+1$
dimensions which are extended in two spatial dimensions to form a wall
structure. Because static wall solutions depend only on one coordinate (the
distance from the wall's core), they can often be found analytically in the
absence of gravity, and perturbatively analytically in its presence. In a
cosmological context, it was soon realised~\cite{ZKO} that the existence of
domain walls with $\eta \, {\scriptstyle \gtrsim} \, 1$~MeV must be ruled out,
because a network of such defects would rapidly evolve to dominate the energy
of the Universe.  Nevertheless, domain walls remain intrinsically interesting
objects to study, for instance for their properties as hypersurfaces; in
cosmology, domain walls have been proposed as a realization of our universe in
higher dimensions~\cite{RubSh}, and are currently being explored as a
possible resolution of the hierarchy problem~\cite{RS}.

When considering `defects', there are two main aspects to understand: their
gravitational (or other particle) interactions and their dynamics. The gravity
of domain walls is an interesting and rather more subtle topic than it might
seem at first sight.  Indeed, unlike all other defects (with the exception of
global strings~\cite{R96}), the wall's metric is not in general
static~\cite{WallMet,IpSi}, but admits a de Sitter-like expansion in its plane.
Moreover, observers experience a repulsion from the wall, and there is a
cosmological horizon at a finite proper distance from the defect's core. This
horizon is a consequence of the choice of coordinates, and in a different set
of coordinates~\cite{BubWa} the wall has the appearance of a bubble which
contracts in from infinite radius to some minimum radius, then re-expands,
undergoing uniform acceleration from the origin. The `horizon' is then simply
the lightcone of the origin in these coordinates, and is somewhat similar to
the horizon of Rindler spacetime.  These results were originally obtained for
infinitesimally thin walls, using the hypersurface formalism developed by
Israel~\cite{Israel}, but can be shown to be robust as an approximate
description of a thick domain wall by a perturbative expansion in the thickness
of the wall~\cite{larryw,GG}, or within the context of a fully nonlinear
treatment of a scalar field coupled to gravity~\cite{FCR}.

The crucial physical difference, then, of the self-gravitating domain wall
spacetime, is the presence of the cosmological horizon, which introduces a
second length scale into the system.  Ordinarily, a defect possesses one length
scale, its thickness $w$, however, the distance to the event horizon of the
domain wall gives another length scale, $u_{\rm h}$, which can be compared to
$w$. These lengths are given in terms of the coupling constants of the theory
and, as taking a thin wall limit turns out to be a very artificial construction
in terms of these underlying parameters, it becomes pertinent to examine
both the gravity and dynamics for a thick domain wall.

The dynamics of topological defects are typically extremely nonlinear, and
their study is usually carried out in the so-called Nambu approximation, where
the full-field theory action of the model in which the 
defects arise is replaced by an approximation based on the 
degrees of freedom of the defect's core. At first
sight, this looks like a perfectly reasonable approximation, since---even for
finite-sized defects, such as string loops and wall bubbles---the thickness
of the defect is typically many orders of magnitude smaller than its size. It
must be noted however, that the Nambu action is only a leading order
approximation of the real action, and is obtained in the limit $\alpha \to 0$,
where $\alpha \propto K w$ is proportional to the defect's typical curvature
and thickness~\cite{MTR,Vanda,RDD,BR}.  However, it is often the case that
the small-scale structure of defects (when $\alpha$ can be significant) is of
particular importance when considering their impact. It seems that quite
generically these are the points at which the defects lose most of their energy
(by Higgs, gravitational or possibly other types of radiation, depending on the
type of defect) and the properties of the small-scale therefore have a direct
impact on the defects' lifespan, which in turn dictates the possible
cosmological implications they can have.

The first attempts to derive effective actions for walls~\cite{Vanda,RDD,BR}
naturally neglected gravity, in the sense that neither curvature of the
background spacetime nor the self-gravity of the wall were considered. The
effect of background spacetime curvature was considered (using the effective
action method discussed below) in~\cite{BBS}, and the self-gravity of the wall
for a very special trajectory was considered in~\cite{RDD}; however, the motion
of a fully self-gravitating thick domain wall (where the curvature of spacetime
is that induced by the wall) has not to date been considered. In this paper, we
address this problem, namely, using the $\lambda \Phi^4$ field theory, we
examine the dynamics of a thick, self-gravitating kink domain wall solution.

There are essentially two methods which have been used to obtain the effective
motion of a thick defect, both of which involve in some way an expansion of the
fields around a well-known solution (such as the hyperbolic tangent kink for a
$\lambda \Phi^4$ wall).  The first approach consists of replacing the solution
in the action and integrating out perpendicular to the defect, which yields an
effective action based on the defect's core. This method has been employed for
instance in~\cite{MTR,RDD,BR,BBS}. The second approach consists of examining
the field equations perturbatively in a relevant parameter (or parameters) such
as $\alpha$, with the equations of motion to a particular order arising as an
integrability condition (see for instance~\cite{RDD}). Here we use this latter
method, adapted to take the wall's self-gravity into account. The reason for
this choice is in fact related to the inclusion of gravity in the problem; the
motion of the core of the defect interacts with massless degrees of freedom in
the bulk (the graviton) and hence a correct application of the effective action
method is less transparent.

The layout of the paper is as follows. In the next section we present the
model, then briefly introduce the Gauss--Codazzi formalism and derive the
Einstein equations in the corresponding ``$3 + 1$'' notation. We end this
section by reviewing our method in the case of a flat background spacetime. In
section~\ref{sec:case1} we solve the field equations for the case $\alpha >
\epsilon$ (where $\epsilon$ characterizes the gravitational interaction of the
scalar field). In section~\ref{sec:sph} we discuss the particular case of a
collapsing spherical wall, and we conclude in the last section.

%
%

\section{The Gauss--Codazzi Formalism}\label{sec:gauss}

Our starting point is the usual Goldstone matter Lagrangian,
\be \label{lag}
  {\cal L} = \left( \nabla_a \Phi \right)^2 - U(\Phi),
\ee
where $\Phi$ is a real Higgs scalar field and $U(\Phi)$ is a
symmetry-breaking potential which we take to be
\be
  U(\Phi) = \lambda \left( \Phi^2 - \eta^2 \right)^2 = 
\lambda \eta^4 V\left ( {\Phi\over\eta}\right ).
\ee
This model admits domain wall solutions, where the Higgs field tends to
different vacua $\Phi = \pm \eta$ at, say, $u = \pm \infty$ for a flat
wall.  This implies the existence of a surface for which $\Phi = 0$, and this
surface defines the defect's core.

As in~\cite{FCR}, we scale out the dimensionful parameters from the
Lagrangian by defining
\be
  X = \Phi / \eta \qquad \qquad \epsilon = 8 \pi G \eta^2;
\ee
$X$ now tends to $\pm 1$ at the vacua. The scalar and Einstein equations are
\bml \label{eom1} \bea
\Box X + \frac 2 {w^2} X \left(X^2 - 1 \right) &=& 0 \\
  {\cal R}_{ab} &=& \epsilon \, \left[ 2 X_{,a} X_{,b} - \frac 1{w^2} \,
                    g_{ab} \left( X^2 - 1 \right)^2 \right], \label{eomR}
\eea \eml
where ${\cal R}_{ab}$ is the spacetime Ricci tensor, and $w$ is proportional to
the inverse mass of the Higgs after the symmetry breaking, and therefore
characterizes the width of topological defects within the theory. Note
that according to~(\ref{eomR}) $\epsilon$ characterizes the gravitational
interaction of the Higgs field.

As mentioned in the introduction, our method for finding the effective dynamics
of domain walls requires the expansion of the quantities appearing in the full
equations of motion in powers of some small parameter. This can be achieved by
splitting these quantities in their components parallel and perpendicular to
the wall's worldvolume $\Sigma$, with the help of the Gauss--Codazzi
formalism~\cite{RDD}. Before we start, a remark on our notation: although we
shall generally use lowercase Latin indices $a, b, \ldots$, we may emphasize
the parallel character of some index by using uppercase Latin letters $A, B,
\ldots$ For instance, the coordinates parallel to the defect will be called
$\sigma^A$. The proper distance from the wall will be denoted by $u$.

A domain wall's core (defined by the location of $X\equiv0$ in the above model)
is a three-dimensional surface in four-dimensional spacetime, and consequently
it admits a (spacelike) unit normal field denoted by $n^a$. This normal field
can be regularly extended off the worldvolume by imposing $n^a \nabla_a n_b =
0$, so that each surface of constant $u$ has a normal field $n^a$, a first
fundamental tensor $h_{ab}$ and a second fundamental tensor $\hat{K}_{ab}$,
the latter two being defined by
\be
\label{I&II}
  h_{ab} = g_{ab} + n_a n_b \qquad \qquad \hat{K}_{ab} = h^c{}_a \nabla_c
           n_b.
\ee
$h_{ab}$ is the projection tensor onto the worldvolume (with our choice of
coordinates, its parallel components are equal to the intrinsic metric's, and
its perpendicular components vanish). $K_{ab}$ is the extrinsic curvature of
$\Sigma$, which lies tangentially to the worldsheet and describes how the wall
curves away from a hyperplane in spacetime. The metric of the wall
spacetime is therefore written in this coordinate system as
\be
  ds^2 = h_{AB} d\sigma^A d\sigma^B - du^2.
\ee

The Gauss and Codazzi integrability conditions for the hypersurface
generated by the wall's core are,
\be \label{gauss}
  \hat{R}_{ab}={\cal R}_{cd}h_a^c h_b^d+\hat{K}_{ac}\hat{K}_b^c-\hat{K} 
    \hat{K}_{ab}+{\cal R}_{acbd}n^c n^d
\ee
\be \label{codazzi}
  \hat{D}_c \hat{K}^c_a - \hat{D}_a \hat{K} = \hat{R}_{de}n^e h_a^d,
\ee
where ${\hat D}_a = h_a^c \nabla_c$. Using these integrability conditions,
the equations of motion~(\ref{eom1}) can be rewritten
\bml \label{gc} \bea
  \frac{\partial h_{ab}}{\partial u} &=& 2  \hat{K}_{ab} \\
  \frac{\partial \hat{K}_{ab}}{\partial u} &=& - \epsilon (X^2-1)^2 h_{ab}
    + \left [ 2 \hat{K}_{ac} \hat{K}^c_b - \hat{K} \hat{K}_{ab} -
    \hat{R}_{ab}
    \right ] + 2 \epsilon  \hat{D}_a X \hat{D}_b X \\
  \frac{\partial^2 X}{\partial u^2} - 2X(X^2-1) &=& 
    - \hat{K} {\partial X \over \partial u} + \hat{D}_a \hat{D}^a X
    \label{gcX} \\
  \frac{\partial \hat{K}}{\partial u} &=& -  \hat{K}_{ab}^2 - \epsilon  
    \left [ 2 X'^2 + (X^2-1)^2 \right ] \label{gcK} \\
  \hat{D}_c \hat{K}^c_a - \hat{D}_a \hat{K} &=& 2 \epsilon  {\partial X
    \over \partial u} \hat{D}_a X \\
  \hat{R} &=& \hat{K}_{ab}^2 - \hat{K}^2 + 2 \epsilon \, \left[ X'^2 - (
X^2 -
    1)^2 + \hat{D}_a X \hat{D}^a X \right ] \label{gcK2} .
\eea \eml
Here we have (without loss of generality) set $w = 1$, which amounts to
using wall rather than Planck units. The mean curvature $\hat{K}$ is the trace
of $\hat{K}_{ab}$ and $\hat{R}_{ab}$ is the worldvolume Ricci tensor. Note
that the equations for $\hat{K}_{ab}$, $\hat{K}$ and the derivatives of $\hat
K$ in the wall correspond respectively to the ``$AB$,'' ``$uu$'' and
non-diagonal Einstein equations. The final equation results from the trace of
(\ref{gauss})
and turns out to be related to the integrability condition which gives the
wall equation of motion.

We now need to identify the two key parameters in these equations, as well as
the dependence of the variables on these parameters. Clearly the gravitational
parameter appears explicitly in the equations, however, the parameter $\alpha$
characterising the motion of the wall is only implicit in the equations. The
first step to identifying this parameter is to quantify what one means by
`motion' of the wall; this is encoded in the components of the extrinsic
curvature on the wall core itself, since this tells us that the wall {\it is}
curved in the ambient spacetime. We therefore set
\be
  \alpha \sim |\hat{K}^a_b(u=0)|.
\ee
Note that this is not a fundamental parameter of the theory, in that it is not
given in terms of any coupling constants or masses, but simply represents the
physical motion of the wall and enables the effect of that motion to be
correctly considered. The extrinsic curvature $\hat{K}_{ab}$, therefore, has
two main contributions: that of the motion of the wall core and that of the
gravity of the wall. This can be estimated by considering the case of the
plane-symmetric wall~\cite{FCR}, for which $\alpha=0$, and
\be \label{PSKab}
\hat{K}_{ab} = {A'}{A} \, {\rm diag} \left( 1, - {\rm e}^{2ct}, 
- {\rm e}^{2ct} \right), 
\ee
where $A(u) = 1 + \epsilon \, A_1(u) + O(\epsilon^2)$, $c = 2\epsilon/3 +
O(\epsilon^2)$. Clearly then, the components of $\hat{K}^a_b$ are
O$(\epsilon)$.

To summarise: in order to describe the motion of a domain wall in a curved
spacetime we naturally have two parameters; one characterising the motion of
the wall itself, $\alpha$, and one the curvature of the ambient spacetime,
$\epsilon$.  The basic procedure for determining the equation of motion of the
wall is to solve (\ref{gc}) order by order in these parameters, investigating
any constraints arising on the extrinsic curvature
at each step. It is perhaps worthwhile briefly reviewing
this process for $\epsilon=0$, since the methodology is very similar when
gravity is included.

We begin by rescaling the extrinsic curvature and parallel coordinates via
\bml\label{wallrescale} \bea
  \hat{K}_{ab} &=& \alpha \, K_{ab} \\
      \sigma^A &=& x^A / \alpha.
\eea \eml

Since we are in flat space the Gauss identity (\ref{gauss}) simplifies to,
\be
  R_{ab} = K_{ac} K^c_b - K K_{ab};
\ee
hence
\be
  K_{ab} = K_{ab}|_0 + \alpha u K_a^c|_0 K_{bc}|_0
\ee
(where ``$|_0$'' indicates that a quantity is evaluated at the wall core) is
actually an implicit exact solution to the $K$-equation. To order $\alpha$, we
see that
\be
  K = K|_0 - \alpha u K^2_{ab}|_0,
\ee
and we can examine the $X$-equation (\ref{gcX}) by setting $X = X + \alpha
X_{_1}$, where $X_0 = \tanh u$, finding
\be \label{case1X1}
  {\cal D} X_{_1} \equiv X_{_1}'' - 2X_{_1}(3X_{_0}^2 -1) = - K|_0
X_{_0}'.
\ee
Ordinarily, we might expect to be able to write the solution $X_{_1}$ in terms
of the basis of eigenfunctions of the operator ${\cal D}$, however, we cannot
do this directly, since $X_{_0}'$ is in fact the zero mode of ${\cal D}$. We
are therefore forced to either deduce that $K|_0=0$, or, we can take the
approach of reference~\cite{BR} and remove the requirement that the equations
of motion be regular at the wall. Since we are looking for freely moving wall
trajectories, we will take the former approach, which can be summed up as an
`integrability requirement': multiplying both sides of (\ref{case1X1}) with
$X_0'$, and integrating over \real{} implies
\be \label{case1X11}
  \left(X_1'X_0'-X_0''X_1\right)_{-\infty}^\infty =\int_{-\infty}^{\infty}
  K_0|_0 (X_0')^2 \,du.
\ee
In order for $X_1$ to have the appropriate asymptotic behavior for large
$u$ it follows that
\be \label{Namb0}
  K_0|_0 = 0 \;\;\; (\Rightarrow \;\; X_{_1}=0).
\ee
Equation~(\ref{Namb0}) is of course the Nambu equation $K|_0 = 0$, which
simply means that the core of the defect, situated at $u=0$, follows the Nambu
dynamics to zeroth order.

We can then repeat this process, expanding order by order to get
\bml\bea
  h_{ab} &=& h_{ab}|_0 + 2\alpha u K_{ab}|_0 + \alpha^2u^2
    K_{ac}|_0 K_{b}^c |_0 \\
  K &=& K|_0 - \alpha u K^2_{ab}|_0 + \alpha^2 u^2 K^a_b|_0 K^b_c|_0
    K^c_a|_0 \\
  X &=& X_0 + \alpha X_1 + \alpha^2 X_2,
\eea\eml
where $ K|_0 = O(\alpha^2)$, and
\be\label{X2fs}
  X_2 = \sech^2 u \int_0^u \cosh^4u \int_{-\infty}^u (u
K_{ab}^2|_0)\sech^4
    u.
\ee
This is sufficient to obtain the leading corrections to the Nambu action via
the integrability constraints to third order. To third order (\ref{gcX}) gives
\be
  [X_3'X_0' - X_3 X_0'']' =  - K_2 X_0^{\prime2}
    = - (K_2|_0 + u^2 K^a_b|_0 K^b_c|_0 K^c_a|_0) X_0^{\prime2},
\ee
hence
\be \label{nam}
K_2|_0 = - \frac{f_2(\infty)}{f_0(\infty)} \, K^a_b|_0 K^b_c|_0 K^c_a|_0
= - \left (\frac{\pi^2}{6} - 1 \right ) \,K^a_b|_0 K^b_c|_0 K^c_a|_0,
\ee
where $ f_n(u) = \int_0^u du \, u^n \sech^4u$.

Note that this process of using the integrability condition to derive a
constraint on the extrinsic curvature uses the $X$-equation to order
O($\alpha^{n+1}$) for a constraint on $K$ to order O($\alpha^n$), and that
this constraint only involves the even part of $K$, since any odd parts
integrate to zero. Moreover, the $K$-equation (\ref{gcK}) shows that the even
part of $K$ to order O($\alpha^n$) depends on the odd part of $K_{ab}^2$ to
order O($\alpha^{n-1}$). Keeping this observation in mind prevents the
unnecessary calculation of corrections to the geometric parameters.

%
%

\section{The Motion of a Wall with Gravity}\label{sec:case1}

In order to include gravity, we will make the initial assumption that gravity
is subdominant to the motion of the wall, i.e.\ $\epsilon < \alpha$. Of course
this need not always be the case, however, the derivation of the wall equations
for $\epsilon > \alpha$ is almost identical to $\epsilon < \alpha$, and an
expansion for general $\epsilon$ and $\alpha$ is so notationally cumbersome
that we choose to present the analysis in this particular case for brevity
and clarity.

First of all, note that the Ricci curvature of the wall is at least of order
O($\alpha^2$), as can be seen from~(\ref{gauss}), since the flat space Ricci
curvature is given in terms of products of the extrinsic curvature, and the
self-gravitating wall has $R_{ab} = O(\epsilon^2)$. We may therefore set
\be 
  {\hat R}_{ab} = \alpha^2 R_{ab}.
\ee
Rescaling the extrinsic curvature and parallel coordinates as
in~(\ref{wallrescale}), and defining
\be \label{deldef}
  \delta = {\epsilon\over\alpha},
\ee
the equations of motion in the presence of gravity (here characterised by
$\delta$) become
\bml \label{gc1} \bea
h'_{ab} &=& 2 \alpha \, K_{ab} \label{gc1h}\\
K'_{ab} &=& - \delta (X^2-1)^2 h_{ab} + \alpha \, \left( 2 K_{ac}
K_{bd} h^{cd} - R_{ab} - K K_{ab} \right) + 2 \delta\alpha^2 \, 
D_a X D_b X \\
X'' &=& 2 X \left(X^2 - 1 \right) - \alpha \, K X' + \alpha^2 D_a D^a
X \label{gc1X}\\
K' &=& - \delta \left[ 2 X'{}^2 + \left(X^2 - 1 \right)^2 \right] -
\alpha \, K_{ab}^2 \label{gc1K}\\
D_c K^c{}_a - D_a K &=& 2 \delta X' D_a X\\
\alpha R &=& \alpha K_{ab}^2 - \alpha K^2 
+ 2 \delta \, \left[ X'^2 - ( X^2 - 1)^2 +
\alpha^2 D_a X D^a X \right ]. \label{gc2K}
\eea \eml

To solve these equations, we expand all quantities with respect to $\alpha$:
\be
  \renewcommand{\arraystretch}{1.2}
  \begin{array}{lclclclcl}
    X      & = & X_0       & + & \alpha \, X_1        & + & \alpha^2 \,
           X_2       & + & \cdots \\
    h_{ab} & = & h_{0\,ab} & + & \alpha \, h_{1\, ab} & + & \alpha^2 \,
           h_{2\,ab} & + & \cdots \\
    K_{ab} & = & K_{0\,ab} & + & \alpha \, K_{1\, ab} & + & \alpha^2 \,
           K_{2\,ab} & + & \cdots \\
    R_{ab} & = & R_{0\,ab} & + & \alpha \, R_{1\, ab} & + & \alpha^2 \,
           R_{2\,ab} & + & \cdots, \\
  \end{array}
\ee
which also implies similar series for the traces $K$ and $R$ with, for
instance,
\be
  K_2 = K_{2\,ab} h_0^{ab} - K_{1\,ab} h_1^{ab} - K_{0\,ab} h_2^{ab} + 
        K_{0\,ab} h_{1\;c}^a h_1^{bc},
\ee
where all indices are raised using $h_{0\,ab}$.  Note that we do not expand in
a double series with $\delta$, since the presence of the $\alpha$ terms in the
RHS of~(\ref{gc1h},c,d) means that at any particular order in $\alpha$ the
series expansion in $\delta$ terminates, as we can see from~(\ref{case10})
below. We can now solve equations~(\ref{gc1}) order by order.

To zeroth order in $\alpha$ we obtain 
\bml \label{case10} \bea
  h_{0\,ab} &=& h_{0\,ab}|_0 \\
  K_{0\,ab} &=& K_{0\,ab}|_0 - \delta f_0(u) \, h_{0\,ab}|_0 \\
        K_0 &=& K_0|_0 - 3 \delta f_0(u) \\
        X_0 &=& \tanh u,
\eea \eml
and (\ref{gc2K}) is identically satisfied for $ X_0 = \tanh u$. We 
define
\bml \bea
  f_n(u) &=& \int_0^u du \, u^n V(X_0) \\
  F_n(u) &=& \int_0^u du \, f_n(u).
\eea \eml
Since $V(X)$ is an even function $f_n$ is odd (respectively, even) for $n$ even
(respectively, odd). As a result $F_n$ is even (respectively, odd) for $n$
even (respectively, odd). By considering the derivative of (\ref{gc2K}) we
can deduce that
\be
  R_0 = R_0|_0 \;\;\; \Rightarrow \;\; R_{_0ab} = R_{_0ab}|_0;
\ee
however, to find the actual value of $R_0|_0$ we need to go to first order
for $X$ (see equation (\ref{R0}) below).

To first order, we immediately obtain
\be
  h_{_1ab} = h_{_1ab}|_0 + 2 u K_{_0ab}|_0 - 2 \delta F_0 h_{_0ab}|_0
\ee
and
\be
  X_{_1}'' - 2X_{_1} (3X_{_0}^2 -1) = - X_{_0}' [ K_{_0}|_0 - 3\delta
f_0].
\ee
The integrability requirement then constrains $K_{_0}|_0 = 0$, and $X_1$
is found to be
\be \label{case1X12}
  X_1 = 3\delta X_0' \int_0^u \frac1{X_0'{}^2} \int_{-\infty}^u f_0 X_0'
      =  - \frac\delta6 \frac{3u+  \tanh u}{\cosh^2 u}.
\ee
Note that $X_1$ is a correction due to the presence of gravity. Then
\bml \bea
  K_1 &=& K_1|_0 - u K_{0\,ab}^2|_0 - 4 \delta X_0' X_1 - \delta^2 G_0 \\
  K_{1\,ab} &=& K_{1\,ab}|_0 - \delta f_0 h_{1\,ab}|_0 + 2 u K_{0\,ac}|_0
    K_0{}^c{}_b|_0 - \delta \left( 2f_1 + F_0 \right) K_{0\,ab}|_0
    \nonumber \\
  & & \phantom{K_{1\,ab}|_0} + \delta G_1(u) h_{0\,ab} - u R_{0\,ab}|_0,
\eea \eml
where the functions $G_0$ and $G_1$ are defined by
\bml \bea
  G_0(u) &=& \int du \left( 4 \tanh u \sech^2u X_1 + 3 f_0^2 \right) \\
  G_1(u) &=& 2 f_0 F_0 + \int du \left( 4 \tanh u \sech^2u X_1 - 3 f_0^2
    \right),
\eea \eml
and are both of odd parity. Finally, from (\ref{gc2K}) one obtains 
\be
\label{R0}
  R_0=K_{0\,ab}^2|_0+2\delta^2 \left[
2X_0'X_1'-4X_0X_1(X_0^2-1)-6f_0^2\right]=
  K_{0\,ab}^2|_0-\frac{8}{3}\delta^2.
\ee

At second order in $\alpha$, the equation for $X_2$ is 
\bea
{\cal D} X_2 &=& 6 X_0 X_1^2 - K_1 X_0' - K_0 X_1'  \\
&=& -K_1|_0X_0'+u K_{0\,ab}^2|_0 X_0' + 
6 X_0 X_1^2 + 4\delta X_1 X_0'{}^2 + \delta^2 G_0
X_0' + 3 \delta X_1' f_0. \nonumber
\eea
Since all but the first term on the RHS have odd parity, the integrability
requirement once again constrains $K_1|_0=0$, and we can solve for $X_2$ 
giving
\be\label{X2cs}
  X_2 = \sech^2 u \int_0^u \cosh^4u \int_{-\infty}^u (u
K_{ab}^2|_0)\sech^4
    u \, du,
\ee
which is an odd function with respect to $u$. Therefore, there are no
corrections to the Nambu equation at first order just as in the case of flat
spacetime.  We do anticipate however that such corrections will appear at
second order, and so proceed to calculate $K_2$. However, as we commented
earlier, the constraints due to the integrability requirement only pertain to
the even part of $K_2$, and this in turn depends on the {\it odd} part of
$K_2'$:
\bea
K_2' &=& - 2 K_0^{ab} K_{1\,ab} + 2 K_{0\,ab} K_{0\,c}{}^b h_1^{ac}
\nonumber \\
& & - 2 \delta \left [ 2 X_0' X_2' + X_1'{}^2 + 2 X_0 X_2 \left( X_0^2 - 1
\right) +  X_1^2 \left( 3X_0^2 - 1 \right) \right ] \\
&=& \phantom - 2 u K_{0\,ab}|_0 \, R_0^{ab}|_0 + [\mbox{even terms}].
\nonumber
\eea
Therefore,
\be
K_2|_{\rm even} = K_2|_0 + u^2 K_{0\,ab}|_0 R_0^{ab}|_0.
\ee

Now, examining the $X$-equation at order O$\left( \alpha^3 \right)$ we find
\be \label{case1X3}
  {\cal D} X_3 = 2 X_1^3 + 12 X_0 X_1 X_2 - K_2 X_0' - K_1 X_1' - K_0
                 X_2'
\ee
and therefore the integrability requirement yields the constraint
\be
  K_2|_0 = - \frac{f_2(\infty)}{f_0(\infty)} \, K_{0\,ab}|_0 \,
R_0^{ab}|_0
    = - \left ( \frac{\pi^2}{6} - 1 \right ) \, K_{0\,ab}|_0 \,
R_0^{ab}|_0.
\ee
which gives us the first perturbation of the Nambu equation. 

To summarize, the first corrections to the Nambu equations of motion (reversing
the rescalings performed) appear at second order in $\alpha$ and are:
\be \label{NamP}
  {\hat K}|_0 = - \left ( \frac{\pi^2}6  - 1 \right )\, {\hat K}_{ab}|_0
\, {\hat R}^{ab}|_0.
\ee

Although we obtained this result assuming
$\epsilon < \alpha$, it is in fact quite general, since in a similar
calculation for $\epsilon > \alpha$ rescaling with respect to $\epsilon$ gives
exactly the same result. Clearly from (\ref{gauss}) this correction has the
correct flat space limit (\ref{nam}), and it 
would seem that the inclusion of gravity
simply modifies the second order correction to the equations of motion, rather
than causing a completely new correction to appear. Indeed the three
dimensional Ricci curvature can be seen in (\ref{gauss}) to relate to the
extrinsic and background geometry of our spacetime.

A totally geodesic ($K_{ab}|_0=0$) solution such as the plane-symmetric wall
always trivially satisfies~(\ref{NamP}). What is of real interest, however, is
the existence of non-totally geodesic solutions verifying the perturbative
equations of motion~(\ref{NamP}). A better understanding of the Nambu
correction term will involve the computation of the three dimensional curvature
tensor $R_{0\,ab}|_0$. This task is undertaken in the next section for the
specific example of a collapsing spherical domain wall.

%
%

\section{The Collapse of a Spherical Domain Wall}\label{sec:sph}

In this section we apply the general equations of motion~(\ref{NamP}) to the
case of a collapsing spherical domain wall. This is perhaps the simplest
non-trivial example of a curved domain wall in curved space-time, i.e. where
both of our perturbation parameters $\alpha$ and $\epsilon$ are not zero. The
spherical domain wall has been already studied in different contexts using
Israel's thin wall formalism~\cite{IpSi,larryw,Maeda}. A first attempt to
study the thick case was undertaken in~\cite{RDD} but only using
equations~(\ref{nam}) valid for a flat space-time background.

Consider, in a spherical system of coordinates $(t,r,\theta,\phi)$, a
non-static scalar field representing a domain wall matter coupled to a
spherically symmetric metric
\be \label{sph.met}
  ds^2 = A^2(t,r) dt^2 - B^2(t,r)dr^2 -r^2 d\Omega_{I\!I}^2.
\ee

The field equations one has to solve are the coupled Einstein and scalar
equations, which can be written in a convenient way as
\bml \label{Fspher} \bea
  \frac{(AB)'}{AB^3} &=& \epsilon r \left( \frac{\dot{X}^2}{A^2}+
    \frac{X'^2}{B^2}  \right) \label{eins1}\\
  \left[ \left( 1-\frac{1}{B^2} \right) r \right]' &=& \epsilon r^2\left(
    \frac{\dot{X}^2}{A^2}+ \frac{X'^2}{B^2} +V(X) \right) \label{eins2}\\
  \frac{\dot{B}}{B}&=& \epsilon r X' \dot{X} \label{eins3}\\
    \frac{A''}{AB^2}-\frac{A'B'}{AB^3}-\frac{\ddot{B}}{BA^2}+\frac{\dot{B}
    \dot{A}}{BA^3}+\frac{2A'}{rAB^2}&=&2\epsilon 
    \frac{\dot{X}^2}{A^2}-\epsilon V(X) \label{int}\\
  \Box X+2X(X^2-1) &=& 0. \label{sca}
\eea \eml
For small values of $\epsilon\ll 1$, consider the field expansion, 
\bml\bea
  X &=& X_0 + \epsilon X_1 + O(\epsilon^2)\\
  A &=& 1 + \epsilon A_1+ O(\epsilon^2)\\
  B &=& 1 + \epsilon B_1+ O(\epsilon^2).
\eea\eml
In order to solve~(\ref{Fspher}) perturbatively one has to first solve the
scalar equation~(\ref{sca}) to zeroth order in $\epsilon$. Then integrating out
the Einstein equations~(\ref{eins1}-\ref{eins3}) we can obtain the first order
$\epsilon$ corrections for the metric $A_1$ and $B_1$. Let us first define the
wall's position and discuss some general features about the solution before
solving~(\ref{Fspher}).

The location of the wall is defined by the zero of the Higgs field, and will
have coordinates $X^a=(t,R(t),\theta,\phi)$. We start by making the observation
that in order for the wall to be identifiably a topological kink solution,
$R(t) \gg 1$, and without loss of generality, we can assume that $X < 0$
(respectively, $X > 0$) for $r < R(t)$ ($r > R(t)$). (Note that these are not
the Gauss--Codazzi gauge coordinates centered on the wall's core.)

We consider the following initial conditions which are compatible with the fact
that, due to its spherical symmetry, the wall is not radiating:
\be
  R(t=0) = R_{\rm i}, \qquad \dot{R}(t=0) = 0.
\ee
The wall's initial bending parameter is defined as $\alpha=1/R_{\rm i}$.

For an inertial observer situated outside the wall, $r > R(t)$, so by
Birkhoff's theorem the exterior metric is Schwarzschild:
\be
  ds^2 = \left(1-\frac{2GM}{r}\right)dt^2-\left(1-\frac{2GM}{r}
         \right)^{-1}dr^2+r^2d\Omega_{I\!I}^2,\qquad r > R(t).
\ee
In the same way, inside the wall the spacetime metric is flat to a very good
approximation. The wall's mass, as measured by a distant observer, is $M = 4
\pi \sigma R_{\rm i}^2$ and the Schwarzchild radius is given by $r_{\rm
S} = 2GM$. In order for $R_{\rm i} > r_{\rm S}$, one must impose the
relation $\alpha>\epsilon$.

Keeping these considerations in mind, let us now proceed with solving our field
equations order by order in $\epsilon$ in the region close to the wall's core,
$r \approx R(t)$. In order to do so, we must also expand the function $R(t)$,
determining the wall's position, in powers of $\epsilon$,
\be
  R(t) = R_0(t) + \epsilon R_1(t) + O(\epsilon^2).
\ee
Of course $R_0(t)$ will in fact be given in terms of a power series in
$\alpha$: $R_0(t)= {\cal R}_0(t) + \alpha^2 {\cal R}_2(t)$ etc.

\begin{figure}
\centerline{\epsfig{file=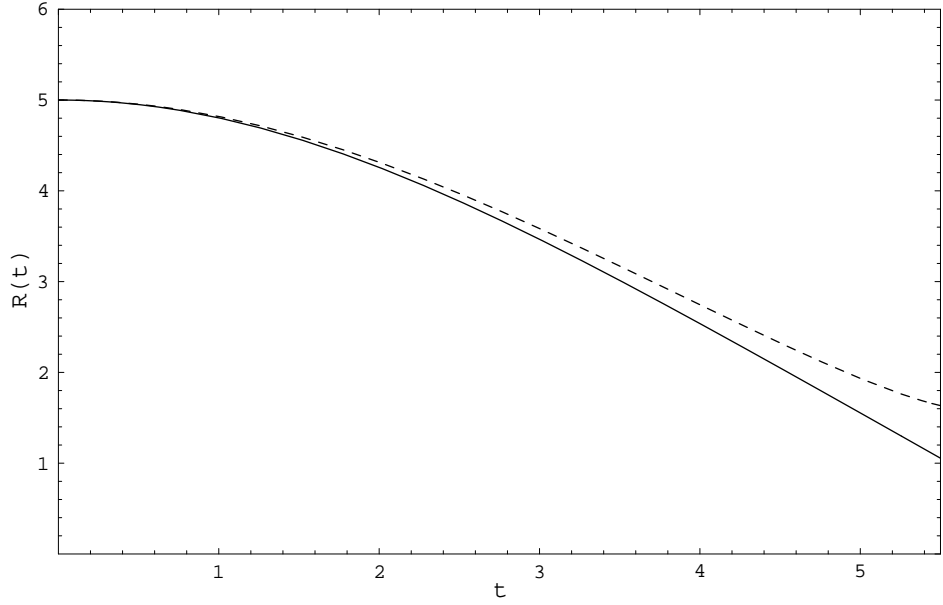,width=12.8cm}}
\caption{The trajectory for the collapsing wall in flat spacetime. 
The solid line is the leading order Nambu trajectory, and the
dashed line the trajectory including O($\alpha^2$) corrections.
For illustrative purposes the rather large value of $\alpha = 1/5$ 
has been chosen.}
\label{fig:mot}
\end{figure}

To zeroth order in $\epsilon$, i.e.\ in a flat background spacetime, we can
define the unit, exterior pointing, normal vector to the wall's core at
$r = R_0(t)$ as
\be \label{normal}
  n^a = (\dot{R_0},1,0,0)/\sqrt{1-\dot{R_0}^2}.
\ee
The intrinsic metric and extrinsic curvature components of the wall (at $r =
R_0(t$)) are easily calculated in turn using~(\ref{I&II}). The extrinsic
curvature for $\epsilon = 0$ is then given by
\be \label{mean}
  K^a_b = \frac{\ddot{R_0}\delta^a_t \delta^t_b}{(1-\dot{R_0}^2)^{3/2}}
        + \frac{\delta^a_\theta \delta^\theta_b + \delta^a_\phi
\delta^\phi_b}
          {R_0 \sqrt{1-\dot{R_0}^2}}
\ee 
and the equation of motion (using (\ref{nam})) is
\be \label{nam1}
\ddot{R_0} = -\frac{2}{R_0}(1-\dot{R_0}^2) - \left (
{\pi^2\over6}-1\right)
\left [ {{\ddot R_0}^3 \over (1 - {\dot R}_0^2)^3} + {2\over
R_0^3} \right] + O(\alpha^3).
\ee
This can be solved iteratively, giving to leading order
\be \label{nam2}
  \left(\frac{{\cal R}_0}{ R_{\rm i}}\right)^4=1-\dot{{\cal R}_0}^2.
\ee
which can be solved analytically and numerically, as can the 
O($\alpha^2$) correction (see~\cite{RDD} and \cite{Let,BKT}), giving 
the wall trajectory as shown in figure~\ref{fig:mot}.
To O($\alpha^2$) we note that $R_0$ satisfies
\be
1 - {\dot R}_0^2 = {R^4_0\over R_{\rm i}^4 } \left [ 1 - {2 C \over R_{\rm
i}^2} \left ( 1 - {R_{\rm i}^6\over R_0^6} \right ) \right ],
\label{secord}
\ee
where $C \equiv f_2(\infty)/f_0(\infty) = \pi^2/6-1$. From~(\ref{nam1})
and~(\ref{nam2}) we note that ${d^n \over dt^n}R_0 = {\rm O}(\alpha^{n-1})$.

In order to solve for the spacetime metric, we need $X_0$. From the work of the
previous section we know that $X_0 = \tanh u + \chi_2$, where $\chi_2$ is given
from equation~(\ref{X2fs}) as
\be
  \chi_2 = {6\over R_0^2(1-{\dot R_0}^2)} \sech^2u \int_0^u \cosh^4 u
           \int_{-\infty}^u u \sech^4 u,
\ee
and is clearly of order O($\alpha^2$). Here, $u$ is the proper distance
from the wall. Using~(\ref{normal}) we note that
\bml\label{ctransfm}\bea
  t &=& t' +\frac{u \dot{R_0}(t')}{\sqrt{1-\dot{R_0}^2(t')}}\\
  r &=& R_0(t')+\frac{u}{\sqrt{1-\dot{R_0}^2(t')}},
\eea\eml
where $(t',u)$ are the coordinates of the point $(t,r)$ in Gaussian Normal
gauge. 

\begin{figure}
\centerline{\epsfig{file=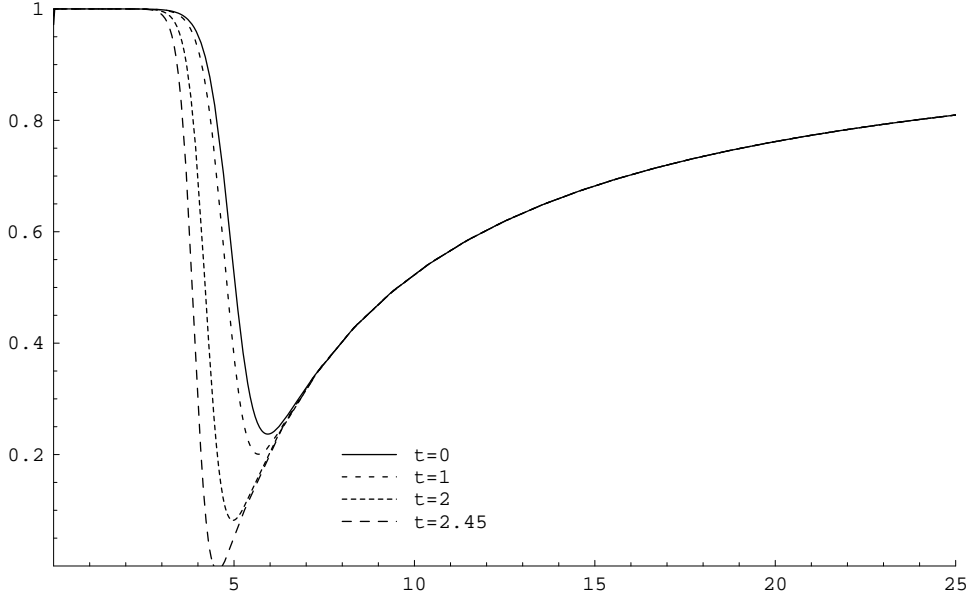,width=12.8cm}}
\caption{The $r-r$ component of the metric for the collapsing wall
solution of figure \ref{fig:mot} for various values of $t$ up to $t_c$,
the critical value of $t$ for which the wall reaches its Schwarzschild radius.
Again, for illustrative purposes, the rather large value of $\epsilon=1/14$
has been chosen.}
\label{fig:mass}
\end{figure}

In order to find the metric, first note that we can directly
integrate~(\ref{eins1}) and~(\ref{eins2}) implicitly, finding
\bml\label{ein1} \bea
  B_1 &=& \frac{1}{2r} \int_0^r r^2 \left [
          \left ( {1+{\dot R}_0^2 (t') \over 1 - {\dot R}_0^2 (t')} \right)
          X_0'^2 + V(X_0) \right ] dr\\
  A_1 &=& -\frac{1}{2r} \int_0^r r^2\left [
          \left ( {1+{\dot R}_0^2 (t') \over 1 - {\dot R}_0^2 (t')} \right)
          X_0'^2 + V(X_0) \right ] dr+ \int_0^r r 
          \left ( {1+{\dot R}_0^2 (t') \over 1 - {\dot R}_0^2 (t')} \right)
          X_0'^2 dr,
\eea \eml
where we have used~(\ref{ctransfm}) to obtain $\dot{u}^2 + u'^2 = [
1+{\dot R}_0^2 (t') ] / [ 1 - {\dot R}_0^2 (t') ]$. We may now 
substitute the form of $X_0(u,t')$ to the required order, 
replace the $r$-integral by a $u$-integral along a line $t 
= \mbox{const}$, then Taylor expand $t'$ around $u = 0$ to the
required order in $\alpha$. For example, $B_1$ gives the mass function
via $B^{-2} = 1 - 2 G M(r) / r \Rightarrow GM(r) = \epsilon r B_1$. 
Computing $B_1$ from the above expression to O($\alpha^2$) yields
\be
GM(r) = {\epsilon R_0^2 \over \sqrt{1 - \dot{R}_0^2}} {\tilde f}_0(u) 
+ 2\epsilon {R_{\rm i}^4\over R_0^3} {\tilde f}_1 (u) 
+ \epsilon R_{\rm i}^2 \sech^2 u \chi_2 (u) + \epsilon {R_{\rm i}^6
\over R_0^6} \left [ {\tilde f}_2(u) + 6 \dot{R}_0^2 u {\tilde
f}_1(u) \right ]
\ee
where we have put
\be
{\tilde f}_n(u) = \int_{-\infty}^u du \, u^n V(X_0).
\ee
This gives the ADM mass (using \ref{secord}) as
\be
GM_{\rm ADM} = \lim_{r\to\infty} \epsilon r B_1 
= {\epsilon R_0^2 \over \sqrt{1 - \dot{R}_0^2}} {\tilde f}_0(\infty) 
+ \epsilon {R_{\rm i}^6\over R_0^6} {\tilde f}_2(\infty)
= \frac43\epsilon \left( R_{\rm i}^2 + \frac{\pi^2} 6 - 1
\right),
\ee
which is indeed constant, and agrees to leading order with the expected
result. Figure~\ref{fig:mass} shows the evolution of the $g_{rr}$ metric
component up to the formation of a black hole.

\begin{figure}[hbtp]
\centerline{\epsfig{file=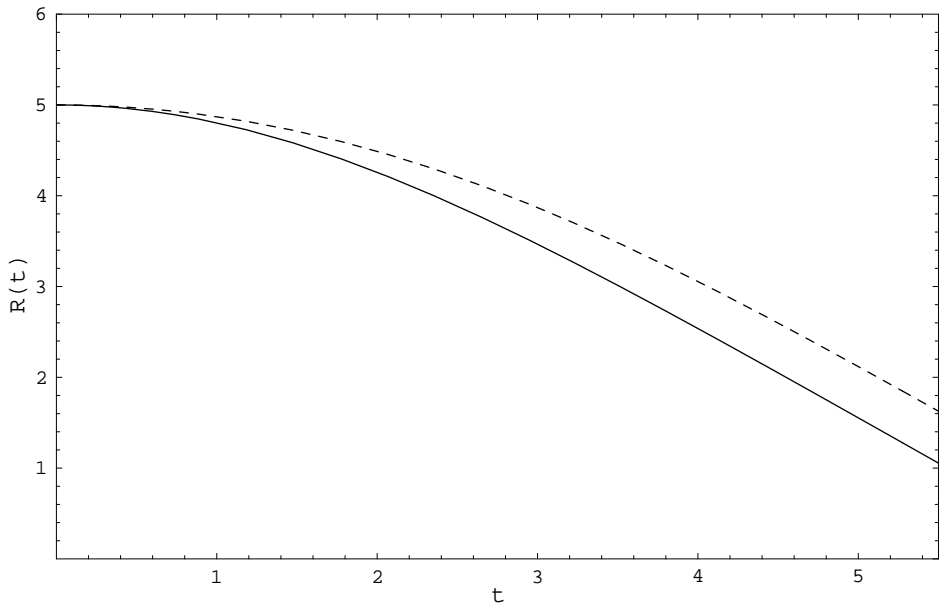,width=12.8cm}}
\caption{A comparison of the leading order (i.e.\ Nambu) flat space 
bubble trajectory (solid line) and the leading order (Nambu) 
trajectory in the presence of gravity. As before,
for illustrative purposes the rather large values of $\alpha = 1/5$
and $\epsilon = 1/14$ have been chosen. Although the bubble actually reaches
its Schwarzschild radius at $t\simeq2.45$ (see fig \ref{fig:mass}), we
have chosen the $t$-range for direct comparison with figure \ref{fig:mot}.}
\label{fig:grav}
\end{figure}

The gravitational Nambu equation for $\epsilon \neq 0$ at $u = 0$ is given by
\be \label{gnam}
  \ddot{R}+\frac{2}{R}nA^2+(A'A+\dot{R}\dot{B}B)n
    +\dot{R}^2\left(\frac{B'}{B}-\frac{A'}{A} \right)
    +\dot{R}\left(\frac{\dot{B}}{B}-\frac{\dot{A}}{A} \right)
    =O(\alpha^2\epsilon),
\ee
where $n = B^{-2} - \dot{R}^2A^{-2}$. Note that the gravitational
correction to $A$ and $B$ is O($\delta$) not O($\epsilon$), 
where $\delta = \epsilon/\alpha$ was defined in~(\ref{deldef}). 
This means that gravitational corrections to the wall motion 
potentially could appear at O($\delta$). Since the flat space wall
equations are intially O($\alpha$), we could have the catastrophic
situation that adding gravity swamps the wall motion, and has a 
superdominant effect. In fact, this turns out not to be the case. 
The computation of the $A$ and $B$
contributions to O($\delta$) in~(\ref{gnam}) shows that they cancel. Instead,
the leading order correction appears at O($\delta\alpha = \epsilon$),
which is therefore subdominant to the flat space motion, and is
plotted in figure \ref{fig:grav}
\be
\label{nambueom}
\ddot{R} = - {2\over R} (1 - \dot{R}^2 ) 
+ 2\epsilon (1 - \dot{R}^2)^{3/2}
+ {2\epsilon (4\ln 2 - 1) \dot{R}^2 \over 3R}
(1 - 6 \dot{R}^2 + 3\dot{R}^4)
\ee
Note that we have included the O($\alpha\epsilon$) correction, since it
is the same order as the finite width correction, however, this has not
been used in the computation of the corrected Nambu trajectory in 
figure \ref{fig:grav}. We should stress that this
correction does not mean that the wall trajectory is no longer Nambu, it
simply gives the right gravitational alteration to the trajectory 
to allow the wall to remain a minimal surface in the now curved 
spacetime. What is important however, is that the correction is 
subdominant to the flat space motion and is
a simple `nudge'; there is no evidence that adding gravity gives any
catastrophic effect which could force a trajectory to be totally geodesic.

What is of real interest to us however, at least in the context of our
general equation (\ref{NamP}), is the correction to the Nambu motion. 
Indeed, having a particular example, we can explicitly calculate $R_{ab}$ 
from Gauss's equation~(\ref{gauss}). Since we have not rescaled our 
quantities here, we can check that the correction term is indeed of the 
right order, and get an idea of the physical implication that the 
correction induces (at least in this particular example). We obtain,
\be
\label{corr}
  \hat{K}_{0}=\frac{\pi^2-6}{\left(1-\dot{R_0}^2\right)^{3/2}}\left(
    \frac{1}{R_0^3}+
\frac{\epsilon}{R_0}\frac{\dot{R_0}^2(1+\dot{R_0}^2)}
{\left(1-\dot{R_0}^2\right)^2}\right).
\ee
The first term is a correction term of order O($\alpha^3$) due to the
bending of the wall, already present in flat space-time as predicted
in~\cite{RDD}. The second correction term is
due to self-gravity appearing at O($\alpha \epsilon$), as we were indeed
expecting (for a spherical wall $\alpha>\epsilon$).  Note 
that the finite width gravitational correction can dominate
the extrinsic curvature correction, depending on the relative magnitudes 
of $\epsilon$ and $\alpha^2$. Finally from the overall 
positive sign we can deduce that the 
corrections to the Nambu motion induce a slowdown of the wall's collapse.

Using (\ref{nambueom}) and (\ref{corr}) we can write the full equation 
of motion (i.e. including the dominant deviation to Nambu motion)
for a collapsing spherical wall up to and including order
$\alpha\epsilon$,
\bea
\ddot{R} = &-& {2\over R} (1 - \dot{R}^2 ) 
+ 2\epsilon (1 - \dot{R}^2)^{3/2}
+ {2\epsilon (4\ln 2 - 1) \dot{R}^2 \over 3R}
(1 - 6 \dot{R}^2 + 3\dot{R}^4)\nonumber \\
&+&(\pi^2-6)\frac{\epsilon}{R}\frac{\dot{R}^2(1+\dot{R}^2)}
{\left(1-\dot{R}^2\right)^2}.
\label{fulleom}
\eea
Before leaving the collapsing wall, it is worthwhile comparing our thick
wall calculation with the results of Ipser and Sikivie, \cite{IpSi}, 
obtained for the collapsing thin wall. In 
order to obtain the Israel thin wall approximation one should reintroduce 
the width parameter $w$ which has been set equal to unity throughout 
this analysis.  Then $\alpha=w/R$ (where R is a typical radius of 
curvature of the wall) is a dimensionless parameter.
The Israel limit is then obtained by letting $\alpha$ and $\epsilon$ tend
to zero, while keeping their quotient $\delta$ fixed. This 
amounts to keeping the horizon distance from the wall finite i.e., 
keeping the wall self-gravitating (as one should in this formalism). 
This also gives us the correct limit without involving the normal 
coordinate $u$ in the limiting procedure.  Taking this limit in
(\ref{fulleom}) gives
\be
\ddot{R} = - {2\over R} (1 - \dot{R}^2 )
+ 2\delta (1 - \dot{R}^2)^{3/2}
\ee
which is obviously only correct to O($\delta$). The translation to the
calculation of Ipser and Sikivie is not direct, since our coordinates
correspond to the interior coordinates of their bubble, however, once the
correct correspondence is made, we do indeed find precise agreement to order
O($\delta$)$=$O($G\sigma$).

%
%

\section{Conclusions}\label{sec:disc}

We have obtained a general equation of motion~(\ref{NamP}) describing a
moving wall in curved spacetime. This was achieved by analytically solving the
Einstein and scalar matter field equations order by order with respect to
two parameters: the wall's bending parameter, $\alpha$, and the gravitational
strength parameter, $\epsilon$, expressing the curving of spacetime. We then
considered a particular example, the collapsing spherical domain wall,
which is perhaps the simplest non-trivial example with both parameters 
different from zero. In the context of this example  we found that 
the corrections to the flat space Nambu motion tended to slow down 
the wall's collapse. 

Throughout this paper we have considered a thick wall with scalar (Higgs)
matter. This was done in order to examine the problem in its analytic
(with respect to the spacetime metric) and most general context and 
also in order to pick up finite width gravitational corrections. As we
discussed at the end of the previous section, in order to obtain the 
Israel thin wall approximation one takes the limit $\alpha\to 0$ 
and $\epsilon\to 0$ keeping $\delta$ fixed. Although the spherical wall
was only explored to O($\delta$), the arguments in section \ref{sec:case1}
show in all generality that the corrections to the  Nambu equation of 
motion $K=0$ are finite width corrections, and hence disappear in
the thin wall limit, giving simply gravitational corrections to 
the flat space trajectory. 

A totally geodesic ($K_{ab}|_0=0$) solution such as the plane-symmetric wall
always trivially satisfies~(\ref{NamP}). With respect to the physical motion of
the wall a totally geodesic solution is trivial and in particular a totally
geodesic wall will not emit gravitational waves. What is of real interest
physically, is the existence of non-totally geodesic solutions such as the
spherical domain wall satisfying~(\ref{NamP}). There have been claims however
(see for the related example of a cosmic string, \cite{christos},
and references therein), that the presence of
gravity in general constrains a defect's core to be totally geodesic. For a
domain wall in particular it has been claimed that the presence of gravity
induces the wall to lose its dynamical degree of freedom and not to 
radiate~\cite{kodama}. We should stress that throughout our treatment this
constraint has not appeared and gravity affects the motion of the defect in a
very natural way in the sense that corrections to the Nambu motion appear at
the same order as in the case of a flat background spacetime.
We suspect that the reason for this discrepancy is that in \cite{kodama}
rather specific asymptotic boundary conditions have been placed on
the spacetime, namely that it asymptote the static planar domain wall 
solution. If, however, a wall is oscillating, we expect that its effective
equation of state will change, analogous to that of the wiggly cosmic string
\cite{vilcart}, which will change the spacetime metric even asymptotically.
Mathematically, this can be seen via the divergent behaviour of the metric
perturbations in the Gaussian Normal gauge due to proper motion of the wall,
and was discussed in the context of higher dimensional domain walls in
\cite{CGR}.
Following \cite{vilcart}, we expect that the equation of state of the wall
will have the form $\sigma T^2 = \sigma_0^3$, or, that perturbatively
the effect on the energy-momentum tensor of the wall will be to increase the
energy by $\delta\sigma$, and decrease the tension by $\delta\sigma/2$. Note
that this perturbation is tracefree, and localised on the wall. In
fact if we regard our domain wall as a $2+1$-dimensional universe, this
energy momentum would be that appropriate to a radiation dominated
cosmology.  Such systems have been explored in a different context in,
for example, \cite{braneu}.

As a final point it should be clear that although we have considered a
four-dimensional spacetime our equations of motion are valid for any
$(n-2)$-brane of an $n$-dimensional spacetime. Our analysis however relies
heavily on the fact that the wall is a hypersurface of the ambient spacetime
i.e.\ that codimension is one. If we were to consider dynamics of strings for
instance, the picture could in principle be quite different.

\section*{Acknowledgements}

We would like to thank Roberto Emparan, David Langlois, Joao
Magueijo and Valery Rubakov for useful discussions.
F.B. is supported by a FAPESP grant; C.C. is supported by PPARC 
and R.G. is supported by the Royal Society.

\def\apj#1 #2 #3.{{\it Astrophys.\ J.\ \bf#1} #2 (#3).}
\def\cmp#1 #2 #3.{{\it Commun.\ Math.\ Phys.\ \bf#1} #2 (#3).}
\def\comnpp#1 #2 #3.{{\it Comm.\ Nucl.\ Part.\ Phys.\  \bf#1} #2 (#3).}
\def\cqg#1 #2 #3.{{\it Class.\ Quant.\ Grav.\ \bf#1} #2 (#3).}
\def\grg#1 #2 #3.{{\it Gen.\ Rel.\ Grav.\ \bf#1} #2 (#3).}
\def\jmp#1 #2 #3.{{\it J.\ Math.\ Phys.\ \bf#1} #2 (#3).}
\def\ijmpd#1 #2 #3.{{\it Int.\ J.\ Mod.\ Phys.\ \bf D#1} #2 (#3).}
\def\mpla#1 #2 #3.{{\it Mod.\ Phys.\ Lett.\ \rm A\bf#1} #2 (#3).}
\def\ncim#1 #2 #3.{{\it Nuovo Cim.\ \bf#1\/} #2 (#3).}
\def\npb#1 #2 #3.{{\it Nucl.\ Phys.\ \rm B\bf#1} #2 (#3).}
\def\phrep#1 #2 #3.{{\it Phys.\ Rep.\ \bf#1\/} #2 (#3).}
\def\pla#1 #2 #3.{{\it Phys.\ Lett.\ \bf#1\/}A #2 (#3).}
\def\plb#1 #2 #3.{{\it Phys.\ Lett.\ \bf#1\/}B #2 (#3).}
\def\pr#1 #2 #3.{{\it Phys.\ Rev.\ \bf#1} #2 (#3).}
\def\prd#1 #2 #3.{{\it Phys.\ Rev.\ \rm D\bf#1} #2 (#3).}
\def\prl#1 #2 #3.{{\it Phys.\ Rev.\ Lett.\ \bf#1} #2 (#3).}
\def\prs#1 #2 #3.{{\it Proc.\ Roy.\ Soc.\ Lond.\ A.\ \bf#1} #2 (#3).}

\end{document}